\documentclass[twocolumn,superscriptaddress]{revtex4}

\usepackage{acronym}
\usepackage{amsmath}
\usepackage{amssymb}
\usepackage{graphicx}

\def\Es{E^\text{s}}
\def\eis{{e_\text{IS}}}
\def\Eint{E^\text{int}}
\def\Eext{E^\text{ext}}
\def\eth{{e_\text{th}}}
\def\qc{{q_c}}
\def\TMC{T_c}
\def\var#1{\text{Var}[#1]}
\def\r{\mathbf{r}}

\begin{document}


\title{Surface tension fluctuations and a new spinodal point in
glass-forming liquids}

\author{C. Cammarota} 
  \affiliation{Dipartimento di Fisica, Universit{\`a} di Roma ``La Sapienza''.}
  \affiliation{Centre for Statistical Mechanics and Complexity (SMC), CNR-INFM.}

\author{A. Cavagna}  
  \affiliation{Centre for Statistical Mechanics and Complexity (SMC), CNR-INFM.}
  \affiliation{Istituto Sistemi Complessi (ISC), CNR, Via dei Taurini 19, 00185 Roma, Italy.}

\author{G. Gradenigo} 
  \affiliation{Dipartimento di Fisica, Universit{\`a} di Trento, via Sommarive 14, 38050 Povo, Trento,
  Italy.}
  \affiliation{INFM CRS-SOFT, c/o Universit\`a di Roma ``La Sapienza'', 00185, Roma, Italy.}

\author{T. S. Grigera} 
  \affiliation{Instituto de Investigaciones Fisicoqu{\'\i}micas
  Te{\'o}ricas y Aplicadas (INIFTA).} 
  \affiliation{Departamento de F{\'\i}sica, Facultad de Ciencias
  Exactas, Universidad Nacional de La Plata, c.c. 16, suc. 4, 1900 La
  Plata, Argentina.}
  \affiliation{CCT La Plata, Consejo Nacional de Investigaciones
  Cient{\'\i}ficas y T{\'e}cnicas, Argentina.}

\author{P. Verrocchio} 
  \affiliation{Dipartimento di Fisica, Universit{\`a} di Trento, via Sommarive 14, 38050 Povo, Trento,
  Italy.}
  \affiliation{INFM CRS-SOFT, c/o Universit\`a di Roma ``La Sapienza'', 00185, Roma, Italy.}
  \affiliation{Instituto de Biocomputaci\'on y F\'{\i}sica de Sistemas
  Complejos (BIFI), Spain.}

\maketitle


\acrodef{IS}{inherent structure}
\acrodef{RFOT}{random first-order theory}


{\bf The dramatic slowdown of glass-forming liquids
  \cite{review:angell88} has been variously linked to increasing
  dynamic \cite{heterogeneities:berthier05, review:ediger00} and
  static \cite{self:prl07, self:nphys08} correlation lengths. Yet,
  empirical evidence is insufficient to decide among competing
  theories \cite{glassthermo:gibbs58, mosaic:kirkpatrick89,
    glassthermo:mezard99, heterogeneities:garrahan02, review:tarjus05,
    glassthermo:moore06}. The \ac{RFOT} \cite{mosaic:kirkpatrick89}
  links the dynamic slowdown to the growth of amorphous static order,
  whose range depends on a balance between configurational entropy and
  surface tension. This last quantity is expected to vanish when the
  temperature surpasses a spinodal point beyond which there are no
  metastable states.  Here we measure for the first time the surface
  tension in a model glass-former, and find that it vanishes at the
  energy separating minima from saddles, demonstrating the existence
  of a spinodal point for amorphous metastable order. Moreover, the
  fluctuations of surface tension become smaller for lower
  temperatures, in quantitative agreement with recent theoretical
  speculation \cite{self:nphys08} that spatial correlations in glassy
  systems relax nonexponentially because of the narrowing of the
  surface tension distribution.}



It is only recently that a \emph{static} correlation length $\xi$ has
been clearly detected in a glassforming liquid
\cite{self:prl07,self:nphys08}, by measuring a point-to-set
correlation function \cite{mosaic:bouchaud04, dynamics:montanari06}.
The idea is to consider a liquid region of size $R$ subject to
amorphous boundary conditions provided by the surrounding liquid
frozen into an equilibrium configuration. The external particles act
as a pinning field favouring internal configurations which best match
the frozen exterior. Clearly, the effect of the border on the
innermost part of the region is smaller as $R$ grows larger. Less
trivially, on lowering the temperature the effect of the amorphous
boundary conditions propagates deeper into the region. More precisely,
if we measure some correlation (or overlap) $q_c(R)$ between the
initial configuration of the region and that reached at infinite time
under the effect of the amorphous boundary conditions, the decay of
$q_c(R)$ is slower the lower $T$ \cite{self:prl07, self:nphys08}. This
demonstrates the existence of an increasing static correlation length
$\xi$.  Irrespective of its precise definition, $q_c(R)$ is the
point-to-set correlation function.

According to \ac{RFOT}, the decay of $q_c(R)$ is regulated by
competition between a surface energy cost, $Y R^\theta$, trying to
keep the region to the same amorphous state as the external
configuration, and a configurational entropy gain, $T \Sigma R^d$,
favouring a transition to another of the exponentially many states,
${\cal N}(R) \sim \exp(R^d \Sigma)$, available to the region
\cite{self:prl07}. The surface tension $Y$ is predicted to vanish for
temperatures higher than a spinodal value. Above this point metastable states
merge into a single ergodic state. The cost/gain terms balance at $R=
\xi\equiv (Y/T\Sigma)^\frac{1}{d-\theta}$: for $R<\xi$ the surface
cost $Y_c R^\theta$ keeps the region in the same state as the external
environment; for $R>\xi$ the entropic gain $T\Sigma R^d$ dominates and
the region is free to rearrange into some other state.  The \ac{RFOT}
length $\xi$ is not only the typical size of the rearranging regions;
it also represents the largest scale over which it is sensible to
define a metastable state: a state defined over a region much larger
than $\xi$ is unstable against fragmentation into sub-regions of
typical scale $\xi$.  For this reason \ac{RFOT} is also called
\emph{mosaic} theory.

The existence of many metastable states is central to \ac{RFOT}. But
for more than one state to exist, a nonzero surface tension is
necessarily required. In fact, one may argue that the surface tension
is as much a fundamental ingredient of the theory as the
configurational entropy, if not more. Hence a direct empirical
investigation of the surface tension is needed in order to put a more
stringent test on \ac{RFOT}. This is what we do here.


Surface tension is defined as the free energy cost per unit area
associated to a surface separating two phases
\cite{review:navascues79}. The determination of the surface free
energy between metastable states is very challenging, first, because
metastable states are amorphous and the interfaces are hard to detect,
and second, because their lifetime is necessarily finite. Excitations
are constantly forming and relaxing: this is the relaxation mechanism
of \ac{RFOT}, through which the whole liquid state is slowly
explored. For our initial study of surface tension we choose to avoid
such complications and focus on \acp{IS}, {\sl i.e.\/} local minima of
the potential energy \cite{landscape:stillinger84}. This approach
measures energy rather than free energy, so we expect our results to
apply quantitatively only at quite low temperatures.

We consider a set of \acp{IS} obtained minimizing equilibrated
instantaneous configurations of a model soft-sphere glassformer (see
Methods). Our lowest working temperature is $T=0.89 \, \TMC$ ($\TMC$ is
the Mode Coupling temperature \cite{rev:goetze_leshouches1987})

\begin{figure}
\includegraphics[angle=0,width=\columnwidth]{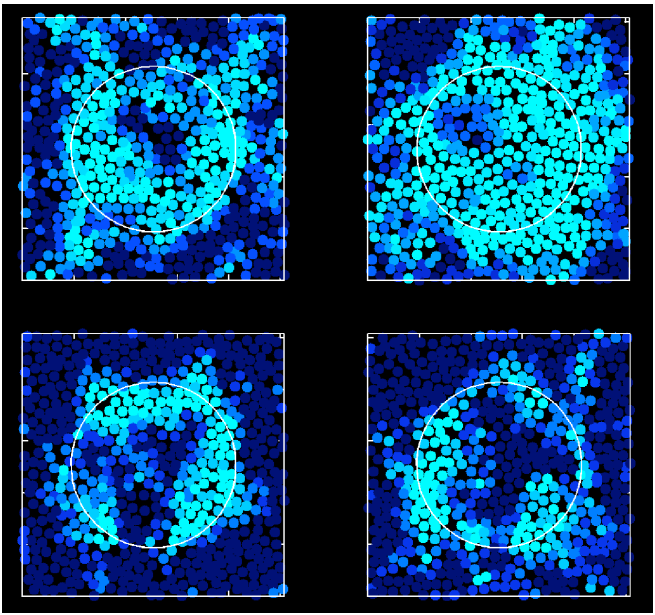}
\caption{{\bf Map of the excitation.} Overlap, i.e.\ similarity,
  between the configuration right after the exchange of the spheres
  and the hybrid inherent structure after minimization. Only a thin
  slice located at half height of the whole three dimensional system
  is shown. Each particle is coloured according to how much it moved
  after the artificial excitation was created (dark blue: small
  displacement, high overlap --- ligth blue: large displacement, small
  overlap).  Upper panel: two configurations at high temperature
  $T=1.33 \, \TMC$. Lower panel: two configurations at low temperature
  $T=0.89 \, \TMC$. The hybrid minimum clearly bears memory (high overlap,
  dark blue) of the parent \acp{IS} far from the boundary of the
  sphere (white circle).  On the other hand, the memory is lower (low
  overlap, light blue) along the interface, where particles have been
  moved the most by the minimization procedure. Although at lower
  temperature the interface is somewhat sharper, it is in general
  quite rough. Strong fluctuations of the overlap along the interface
  are evident, clearly indicating that there are large surface tension
  fluctuations.  Note that the hybrid configuration is nevertheless a
  typical IS of the system: the only reason why we can visualize the
  interface is that we know \emph{a priori} the shape and position of
  the excitation and we can use the parent configurations as reference
  to calculate overlap. Without this information, it would be
  impossible to distiguish the hybrid inherent structure from any
  other one.}
\label{fig:cool}
\end{figure}

Given a pair $\alpha$ and $\beta$ of ISs, we exchange between them all
particles located within a sphere of radius $R$.  We then minimize the
two configurations thus obtained to produce two new \acp{IS}. Each of
them is a hybrid minimum, resembling the ``parent'' minima far from
the surface of the sphere but rearranged close to it
(Fig.~\ref{fig:cool}).

\begin{figure}
\includegraphics[angle=-90,width=\columnwidth]{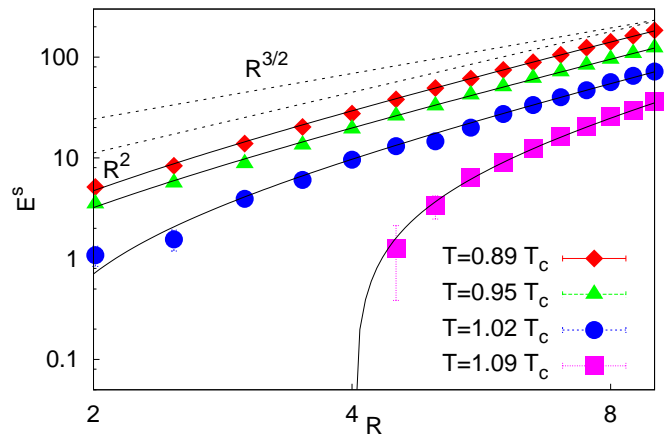}
\caption{{\bf Surface energy.} $\Es$ as a function of the
  excitation radius $R$ at various temperatures.
  A single power law fit $R^\theta$ would give an exponent $\theta
  >2$, which is hard to justify.  On the other hand, especially at
  high $T$ it is clear that there are small $R$ corrections to a
  single power law. Full lines are fits using
  equation~\eqref{eq:UpsilonR}, with $\theta=2$ and
  $\omega=1.5$. Dotted lines show as a reference the power laws with
  $\theta=2$ and $\theta=3/2$. Error bars are smaller than symbol
  size. }
\label{EvsR}
\end{figure}

For each hybrid \ac{IS} we compute the surface energy $\Es_{\alpha
  \beta}$ by subtracting the parent \acp{IS}' contribution (see
Methods).  Fig.~\ref{EvsR} shows the sample-averaged surface energy,
$\Es$, vs.\ $R$ for several temperatures. There is a well-defined
relationship between surface energy and size, regulated by the
temperature: at fixed $R$, $\Es$ increases by decreasing $T$. The data
do not correspond to a single power-law scaling, so we propose
\begin{equation}
  \Es = Y_\infty R^\theta - \delta R^\omega ,
 \label{eq:UpsilonR}
\end{equation}
where $Y_\infty$ is the asymptotic surface tension and $\omega <
\theta$. The subleading $\delta R^\omega$ correction is quite natural.
It is present in liquids (with $\omega=1$) due to curvature effects
\cite{review:navascues79}, and in disordered systems, where it may
arise either from bulk contributions, as in the random field Ising
model \cite{tension:imry75}, or from interface roughening that lowers
the surface energy, as in the random bond Ising model
\cite{tension:halpin95}.

We must choose the exponents of eq.~\eqref{eq:UpsilonR} with some
criterion, because the nonlinear fit with four parameters is
marginally stable, and many sets of parameters give good fits. Our
data strongly suggest the conservative choice $\theta=2$, which seems
to describe the large $R$ behaviour better than the alternative
$\theta=3/2$ predicted by a wetting argument
\cite{mosaic:xia00,mosaic:dzero05} . The value $\theta=2$ is also
found in spin models with finite range interactions
\cite{nucleation:franz05}.  To fix $\omega$, we take
eq.~\eqref{eq:UpsilonR} as valid for the whole population of surface
energies (instead of just the average), and ascribe all fluctuations
of $\Es_{\alpha\beta}$ to the quantity
\begin{equation}
  \delta_{\alpha\beta} = \frac{Y_\infty R^2 - \Es_{\alpha\beta}
  } {R^{\omega}} . \label{eq:deltaab}
\end{equation}
We then require that the variance of $\delta_{\alpha\beta}$ be
independent of $R$, which is the typical behaviour of random systems
\cite{tension:huse85, tension:halpin95}. This procedure (see Methods)
gives $\omega =1.5(2)$.

\begin{figure}
\includegraphics[angle=-90,width=\columnwidth]{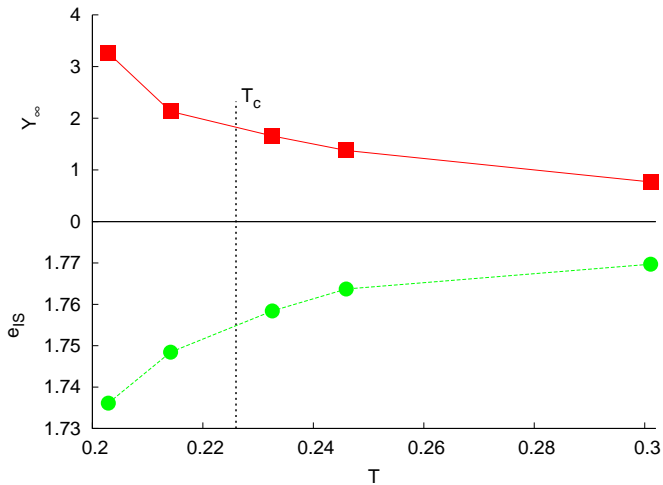}
\caption{{\bf Surface tension vs.\ temperature.} Upper panel:
  $Y_\infty$ as a function of the quenching temperature of the
  inherent structures. The vertical dotted line marks the mode
  coupling temperature.  The surface tension decreases on increasing
  $T$, although too smoothly to indicate a sharp spinodal
  temperature. Lower panel: inherent structure energy as a function of
  the quenching temperature.  A comparison of the two curves clearly
  shows that they are correlated.  }
\label{YvsT}
\end{figure}

With $\theta$ and $\omega$ thus fixed, eq.~(\ref{eq:UpsilonR}) fits
the $\Es(R)$ data very well (Fig.~\ref{EvsR}), and we obtain the
asymptotic surface tension $Y_\infty$ as a function of $T$
(Fig.~\ref{YvsT}, top). We find that $Y_\infty$ decreases for higher
$T$, and becomes quite small above $\TMC$. Yet, the decrease is rather
smooth, so that it is hard to define a spinodal temperature.

\begin{figure}
\includegraphics[angle=-90,width=\columnwidth]{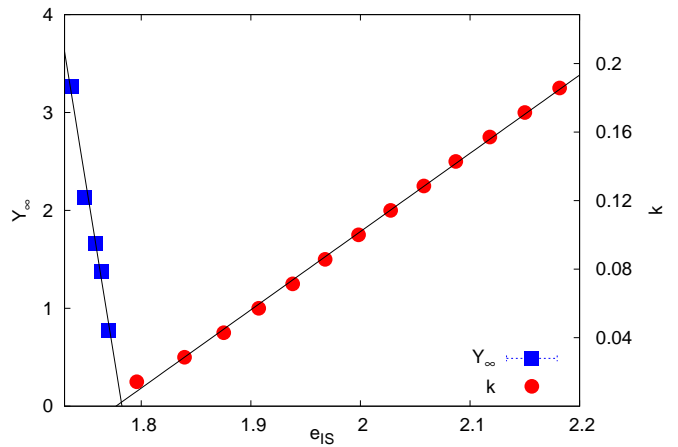}
\caption{{\bf The spinodal point.} Left: $Y_\infty$ (squares) vs.\
  \ac{IS} energy.  Right: intensive saddle instability index $k$
  (circles) vs. \ac{IS} energy.  Lines are linear fits to the
  data. Both the surface tension and the instability index seem to
  vanish at a similar energy, the threshold $\eth$, which is therefore
  the spinodal point.  Instability index data are from
  ref.~\onlinecite{self:jcp06}. }
\label{YKvseis}
\end{figure}

On the other hand, the plot of the average \ac{IS} energy, $\eis$,
vs.\ $T$ (Fig. 3, bottom) suggests using the \emph{energy} as a
control parameter. Indeed, the $Y_\infty$ vs.\ $\eis(T$) curve is
nearly linear (Fig.~\ref{YKvseis}, left), indicating that $Y_\infty$
vanishes quite sharply at a well-defined energy $\eth$. Interestingly
enough, $\eth$ is very close to the \emph{threshold}, {\sl i.e.\/} the
energy below which minima start to dominate the energy landscape
\cite{self:prl02, self:jcp06}. The threshold $\eth$ is defined as the
point where the instability index of saddles vanishes
(Fig.~\ref{YKvseis}, right).  Hence $\eth$ is the true spinodal point
of amorphous order, fixing the upper limit of stability of the
\ac{RFOT} mechanism.

\begin{figure}
\includegraphics[angle=-90,width=\columnwidth]{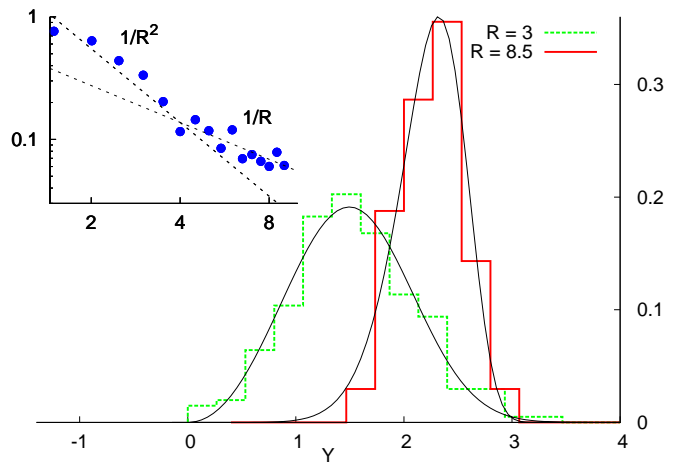}
\caption{{\bf Surface tension distribution.} Normalized histograms of
  the surface tension for $T= 0.89 \, \TMC$ at two values of $R$. Note
  that this distribution does not depend on the exponent $\omega$ and
  that its average tends to $Y_\infty(T)$ for $R\to\infty$.  The
  distribution narrows on increasing $R$, and thus on lowering the
  temperature. Solid lines correspond to fits with the Weibull distribution
  $P(Y) = -d/dY \exp[-(Y/y_{c})^{\hat\zeta}]$
  (see Ref.~\onlinecite{self:nphys08}). Inset: variance of
  $Y_{\alpha \beta}$ vs.\ $R$. For small $R$ this quantity seems to
  decay as $1/R^2$, but at larger $R$ it is clearly larger. Note that
  that, $\var{Y_{\alpha \beta}}= \var{Y_\infty} +
  \var{\delta_{\alpha\beta}}/R^{4-2\omega}$.  If
  $\var{\delta_{\alpha\beta}}$ does not depend on $R$ and $\omega \sim
  3/2$, then $\var{\delta_{\alpha\beta}}/R^{4-2\omega} \sim1/R$, which
  is compatible with the large $R$ behaviour of the data.}
\label{fig:Ydist}
\end{figure}

We expect the surface tension to fluctuate when considering different
pairs of \acp{IS}. We thus compute the distribution of the
single-sample surface tension,
\begin{equation}
  Y_{\alpha \beta}\equiv \frac{\Es_{\alpha \beta}}{R^2} .
\end{equation}
The distribution $P(Y,R,T)$ for two values of $R$ is shown in
Fig.~\ref{fig:Ydist}. The first thing we notice is that the
distribution is quite broad.  This means that the original \ac{RFOT},
which assumed a sharp value of $Y$, cannot hold strictly. If there is
a single surface tension, a region smaller than $\xi$ cannot
rearrange, so that a nonfluctuating surface tension implies a sharp
drop of the point-to-set correlation $q_c(R)$ at $R\sim \xi$.  With a
fluctuating surface tension, on the other hand, \emph{any} region can
decorrelate, as long as there are target states with surface tension
$Y < T\Sigma R^{d-\theta}$. Our finding is thus consistent with the
smooth decay of the correlation observed numerically
\cite{self:prl07,self:nphys08}.

The second important result is that the distribution narrows as $R$
grows. To interpret this correctly, note that $P(Y)$ depends on $T$
both directly through the quench temperature of the \acp{IS} and
indirectly through $R$. Since the only relevant lengthscale is
$\xi(T)$, the typical size of the rearranging regions, we can define
\begin{equation}
P(Y,T) \equiv P(Y, R=\xi(T), T),
\label{eq:scale}
\end{equation}
where $\xi(T)$ is taken from simulations \cite{self:nphys08}. When $T$
decreases, $\xi(T)$ grows, $P(Y,T)$ narrows, and the decay of
$q_c(R,T)$ sharpens. This is indeed the numerical finding
\cite{self:nphys08}: the decay is well described by a compressed
exponential,
\begin{equation} 
\qc(R,T) \sim \exp\left[{-(R/\xi)}^{\zeta}\right],
\label{anomaly}
\end{equation}
where the anomaly $\zeta$ grows from $1$ at high temperature up to
$\sim 4$ at low $T$ (higher $\zeta$ means sharper decay).  The link
between $P(Y,T)$ and sharpness of decay can be made more quantitative.
The overlap $\qc(R)$ is given by \cite{self:nphys08}
\begin{equation}
  \label{eq:qc}
  \qc(R,T) = \int_{T\Sigma R^{d-\theta}}^\infty P(Y,T;R) \,dY.
\end{equation}
Using the approximation \eqref{eq:scale} (see Methods) and defining
$y=T\Sigma R^{d-\theta}$ we obtain
\begin{equation}
 \hat{q}_c(y,T) = \int_y^\infty \!\!\!\! P(Y,T) \, dY.
 \label{crudo}
\end{equation}
In this way we avoid the complexity $\Sigma$, whose estimate is always
very tricky \cite{self:coluzzi99}.  If eq.~\eqref{anomaly} holds, we
must have a similar compressed exponential form, $\hat{q}_c \sim \exp
[ -(y/y_c)^{\hat\zeta}]$, with the two anomalies related by $\hat\zeta
= \zeta/(d-\theta)$. We computed $\hat{q}_c(y,T)$ at the lowest
temperature, $T=0.89 \, \TMC$, where the the difference between energy and
free energy should be less harmful. A compressed exponential fit gives
$\zeta \approx 4.2(1)$, encouragingly close to the value
$\zeta=4.0(6)$ of ref.~\onlinecite{self:nphys08}. The agreement on the
values of $\zeta$ found with two completely different protocols puts
the generalized \ac{RFOT} on a firmer basis.


The present study provides independent evidence of the existence of a
surface tension distribution with the properties required by a
generalized \ac{RFOT}, and it thus supports its validity as a
description of deeply supercooled liquid.  The fact that $P(Y,T)$
broadens and clusters around $Y\sim 0$ for $T\gtrsim \TMC$, (i.e.\ for
$E \geq E_{th}$) is significant in two ways. First, it implies a
crossover from many states to one state through a spinodal mechanism:
a null surface energy cost means that rearrangements are no longer
excitations. Secondly, it says that this transition is smooth: instead
of disappearing abruptly at $\TMC$ (as in mean-field
\cite{self:castellani05}) metastable states slowly ``fade out'',
because excitations are becoming less and less costly on average and
because states are slowly merging with each other, as indicated by
many pairs of states having near-zero surface tension.


\section*{Methods}

\paragraph{System.} 

We have simulated a three dimensional soft-sphere binary mixture
\cite{soft-spheres:bernu87} using Metropolis Monte Carlo with particle
swaps \cite{self:pre01}. The pair potential is
$v_{ij}(\r_i,\r_j)=[(\sigma_i+\sigma_j)/|\r_i-\r_j|]^{12}$, with a
polynomial smooth cut-off at large distances (parameters as given in
Ref.~\onlinecite{self:nphys08}). The mode-coupling temperature for
this system is $\TMC=$0.226 \cite{soft-spheres:roux89}. A system of
$N=16384$ particles was considered in a box of lenght $25.4$ (in such
units the number density is $1$).

\paragraph{Minimization.} 

Minimization of instantaneous configurations was done with the LBFGS
algorithm \cite{algorithm:liu89}. After exchanging the particles of
the sphere, some pairs of particles end up at very short distances,
giving very high energies and gradients that tend to destabilize the
minimizer. For these configurations, we have minimized with a
combination of $100$ standard Metropolis Monte Carlo steps at $T=0$
plus LBFGS. The sphere is placed so that the density and composition
of the resulting configurations are identical to those of the initial
\acp{IS}.

\paragraph{Definition of surface energy.}

We indicate with $E_{\alpha\beta}$ and $E_{\beta\alpha}$ the energies
of the hybrid configurations. The first label indicates the original
parent configuration outside the sphere.  The surface energy is
\begin{align}
  \Es_{\alpha\beta} &= E_{\alpha\beta} - \Eint_\beta - \Eext_\alpha, \\
  E^{\substack{ \text{ext} \\ \text{int}}} &= \sum_{i,j: |\r_i| \gtrless
    R} v_{ij}(\r_i-\r_j) \ .
\end{align}

\paragraph{The exponent $\omega$.} 

Our estimation of $\omega$ relies on the physical hypothesis that the
prefactor $\delta$ in the correction term of Eq.~(\ref{eq:UpsilonR})
(unlike $Y_{\infty}$) has large fluctuations even for $R\to\infty$.
This behaviour is rather ubiquitous in disordered media
\cite{tension:halpin95, tension:huse85,
  tension:imry75} and finds some confirmation here by the violation of
the central limit theorem shown by the variance of $Y_{\alpha \beta}$
at large $R$ (see Fig.~\ref{fig:Ydist}, inset). We fix $\omega$ at the
value for which the variance $\var{\delta(\omega)}$ is
independent of$R$ with the following self-consistent procedure: for a
running value of the exponent, $\tilde{\omega}$, we fit the average of
$E_s$ via Eq.~(\ref{eq:UpsilonR}) to obtain $Y_\infty(\tilde{\omega})$
and a variance $\var{\delta(\tilde\omega)}$. At large $R$
\begin{equation}
  \var{E_s} \sim \var{\delta(\tilde{\omega})}  R^{2
     \tilde{\omega}} ,
\label{VarE}
\end{equation}
where $\var{\tilde{\omega}}$ will in general depend on $R$. A similar
formula holds for $\tilde\omega = \omega$, but with $R$-indepdent
variance, so that
\begin{equation}
  \log \var{\delta(\tilde\omega)} = \log \var{\delta(\omega)}
  + 2(\omega-\tilde\omega) \log R. 
\label{logVardi}
\end{equation}
The procedure is to fit $\log\var{\delta}$ vs.\ $\log R$ for several
values of $\tilde\omega$ to obtain a slope $a(\tilde\omega)$ which is
finally fitted to $a(\tilde\omega)=2(\omega-\tilde\omega)$ to obtain
$\omega$. When we do this we do not find any trend of $\omega$ with
the temperature. Thus, to get rid of unwanted thermal noise in the
determination of this exponent we use the lowest $T$ to fix its
value. This procedure gives $\omega = 1.5(2)$, which is very much
within the range of values provided by nonlinear fit of the data
surface energy with all four parameters free.

\paragraph{Calculation of the anomaly.}

The distribution $P(Y; R)$ is peak-shaped in $Y$; as $R$ is increased
the peak narrows and shifts to the right, approaching
$Y_\infty$. Eq.~(\ref{eq:qc}) says that the overlap is the area of
$P(Y;R)$ to the right of $T\Sigma R$. Around $\xi$, $\qc(R)$ is
rapidly decaying, so that $T\Sigma R$ must be greater than the peak
position for all $R>\xi$, and the integral is just measuring the area
of the right tail of $P(Y;R)$. To study the shape of $\qc(R)$ near the
inflection point $\xi$ (which dominates the value of $\zeta$), we may
thus approximate $P(Y; R)$ with $P(Y; \xi)$ (with the consequence of
slightly overestimating the sharpness of the decay).

\paragraph{The minimal energy with frozen environment.}

We have treated the external and the internal parts of the spherical
excitations symmetrically. However, one may argue that when a region
rearranges, it chooses the target state with the {\it minimum} energy
with respect to the external environment (S.~Franz, private
communication).  This is certainly the case in the numerical
experiments of ref.~\onlinecite{self:nphys08}, where the amorphous
boundary of the rearranging region is really frozen.  We therefore
define the minimal surface energy at fixed external IS,
\begin{equation}
  \Es_{\beta_0,\alpha} = \min_\beta \{ E_{\alpha\beta} \} -
  \Eext_\alpha - \Eint_{\beta_0} , 
\end{equation}
where $\beta_0$ is the $\beta$ that minimizes $E_{\alpha\beta}$. From
this minimum surface energy at fixed $\alpha$, we can obtain the
distribution $P_0(Y)$ by varying the external IS $\alpha$.  This
$P_0(Y)$ is the distribution that must be plugged into the generalized
\ac{RFOT} expression for the overlap, \eqref{eq:qc}, to get $\zeta$.
The value of the anomaly obtained in this way is $\zeta \approx 4.4$,
comparable to the value $\zeta\approx 4.2$ found with $P(Y)$. Indeed,
apart form the poorer statistics that makes $P_0(Y)$ considerably more
noisy than $P(Y)$, the form of the two distributions is very similar.

\acknowledgements

We thank G.~Biroli, J.-P.~Bouchaud, S.~Franz, I.~Giardina and
F.~Zamponi for several important remarks, and ECT* and CINECA for
computer time. The work of TSG was supported in part by grants from
ANPCyT, CONICET, and UNLP (Argentina).


\bibliographystyle{naturemag}

\bibliography{riokurdo}

\begin{thebibliography}{10}
\expandafter\ifx\csname url\endcsname\relax
  \def\url#1{\texttt{#1}}\fi
\expandafter\ifx\csname urlprefix\endcsname\relax\def\urlprefix{URL }\fi
\providecommand{\bibinfo}[2]{#2}
\providecommand{\eprint}[2][]{\url{#2}}

\bibitem{review:angell88}
\bibinfo{author}{Angell, C.}
\newblock \bibinfo{title}{Perspective on the glass transition}.
\newblock \emph{\bibinfo{journal}{J. Phys. Chem. Solids}}
  \textbf{\bibinfo{volume}{49}}, \bibinfo{pages}{863} (\bibinfo{year}{1988}).

\bibitem{heterogeneities:berthier05}
\bibinfo{author}{Berthier, L.} \emph{et~al.}
\newblock \bibinfo{title}{Direct experimental evidence of a growing length
  scale accompanying the glass transition}.
\newblock \emph{\bibinfo{journal}{Science}} \textbf{\bibinfo{volume}{310}},
  \bibinfo{pages}{1797--1800} (\bibinfo{year}{2005}).

\bibitem{review:ediger00}
\bibinfo{author}{Ediger, M.~D.}
\newblock \bibinfo{title}{Spatially heterogeneous dynamics in supercooled
  liquids}.
\newblock \emph{\bibinfo{journal}{Annu. Rev. Phys. Chem.}}
  \textbf{\bibinfo{volume}{51}}, \bibinfo{pages}{99--128}
  (\bibinfo{year}{2000}).

\bibitem{self:prl07}
\bibinfo{author}{Cavagna, A.}, \bibinfo{author}{Grigera, T.~S.} \&
  \bibinfo{author}{Verrocchio, P.}
\newblock \bibinfo{title}{Mosaic multistate scenario versus one-state
  description of supercooled liquids}.
\newblock \emph{\bibinfo{journal}{Physical Review Letters}}
  \textbf{\bibinfo{volume}{98}}, \bibinfo{pages}{187801}
  (\bibinfo{year}{2007}).
\newblock \urlprefix\url{http://link.aps.org/abstract/PRL/v98/e187801}.

\bibitem{self:nphys08}
\bibinfo{author}{Biroli, G.}, \bibinfo{author}{Bouchaud, J.-P.},
  \bibinfo{author}{Cavagna, A.}, \bibinfo{author}{Grigera, T.~S.} \&
  \bibinfo{author}{Verrocchio, P.}
\newblock \bibinfo{title}{Thermodynamic signature of growing amorphous order in
  glass-forming liquids}.
\newblock \emph{\bibinfo{journal}{Nature Phys.}} \textbf{\bibinfo{volume}{4}},
  \bibinfo{pages}{771--775} (\bibinfo{year}{2008}).

\bibitem{glassthermo:gibbs58}
\bibinfo{author}{Gibbs, J.~H.} \& \bibinfo{author}{DiMarzio, E.~A.}
\newblock \bibinfo{title}{Nature of the glass transition and the glassy state}.
\newblock \emph{\bibinfo{journal}{J. Chem. Phys.}}
  \textbf{\bibinfo{volume}{28}}, \bibinfo{pages}{373--383}
  (\bibinfo{year}{1958}).
\newblock \urlprefix\url{http://link.aip.org/link/?JCP/28/373/1}.

\bibitem{mosaic:kirkpatrick89}
\bibinfo{author}{Kirkpatrick, T.~R.}, \bibinfo{author}{Thirumalai, D.} \&
  \bibinfo{author}{Wolynes, P.~G.}
\newblock \bibinfo{title}{Scaling concepts for the dynamics of viscous liquids
  near an ideal glassy state}.
\newblock \emph{\bibinfo{journal}{Phys. Rev. A}} \textbf{\bibinfo{volume}{40}},
  \bibinfo{pages}{1045--1054} (\bibinfo{year}{1989}).

\bibitem{glassthermo:mezard99}
\bibinfo{author}{M\'ezard, M.} \& \bibinfo{author}{Parisi, G.}
\newblock \bibinfo{title}{Thermodynamics of glasses: A first principles
  computation}.
\newblock \emph{\bibinfo{journal}{Phys. Rev. Lett.}}
  \textbf{\bibinfo{volume}{82}}, \bibinfo{pages}{747--750}
  (\bibinfo{year}{1999}).

\bibitem{heterogeneities:garrahan02}
\bibinfo{author}{Garrahan, J.~P.} \& \bibinfo{author}{Chandler, D.}
\newblock \bibinfo{title}{Geometrical explanation and scaling of dynamical
  heterogeneities in glass forming systems}.
\newblock \emph{\bibinfo{journal}{Phys. Rev. Lett.}}
  \textbf{\bibinfo{volume}{89}}, \bibinfo{pages}{035704}
  (\bibinfo{year}{2002}).

\bibitem{review:tarjus05}
\bibinfo{author}{Tarjus, G.}, \bibinfo{author}{Kivelson, S.~A.},
  \bibinfo{author}{Nussinov, Z.} \& \bibinfo{author}{Viot, P.}
\newblock \bibinfo{title}{The frustration-based approach of supercooled liquids
  and the glass transition: a review and critical assessment}.
\newblock \emph{\bibinfo{journal}{J. Phys.: Condens. Matter}}
  \textbf{\bibinfo{volume}{17}}, \bibinfo{pages}{R1143} (\bibinfo{year}{2005}).

\bibitem{glassthermo:moore06}
\bibinfo{author}{Moore, M.~A.} \& \bibinfo{author}{Yeo, J.}
\newblock \bibinfo{title}{Thermodynamic glass transition in finite dimensions}.
\newblock \emph{\bibinfo{journal}{Phys. Rev. Lett.}}
  \textbf{\bibinfo{volume}{96}}, \bibinfo{pages}{095701}
  (\bibinfo{year}{2006}).
\newblock \urlprefix\url{http://link.aps.org/abstract/PRL/v96/e095701}.

\bibitem{mosaic:bouchaud04}
\bibinfo{author}{Bouchaud, J.-P.} \& \bibinfo{author}{Biroli, G.}
\newblock \bibinfo{title}{On the
  {A}dam-{G}ibbs-{K}irkpatrick-{T}hirumalai-{W}olynes scenario for the
  viscosity increase in glasses}.
\newblock \emph{\bibinfo{journal}{J. Chem. Phys.}}
  \textbf{\bibinfo{volume}{121}}, \bibinfo{pages}{7347--7354}
  (\bibinfo{year}{2004}).
\newblock \urlprefix\url{http://link.aip.org/link/?JCP/121/7347/1}.

\bibitem{dynamics:montanari06}
\bibinfo{author}{Montanari, A.} \& \bibinfo{author}{Semerjian, G.}
\newblock \bibinfo{title}{Rigorous inequalities between length and time scales
  in glassy systems}.
\newblock \emph{\bibinfo{journal}{J. Stat. Phys.}}
  \textbf{\bibinfo{volume}{125}}, \bibinfo{pages}{23--54}
  (\bibinfo{year}{2006}).

\bibitem{review:navascues79}
\bibinfo{author}{Navascues, G.}
\newblock \bibinfo{title}{Liquid surfaces: theory of surface tension}.
\newblock \emph{\bibinfo{journal}{Rep. Progr. Phys.}}
  \textbf{\bibinfo{volume}{42}}, \bibinfo{pages}{1131--1186}
  (\bibinfo{year}{1979}).
\newblock \urlprefix\url{http://stacks.iop.org/0034-4885/42/1131}.

\bibitem{landscape:stillinger84}
\bibinfo{author}{Stillinger, F.~H.} \& \bibinfo{author}{Weber, T.~A.}
\newblock \bibinfo{title}{Packing structures and transitions in liquids and
  solids}.
\newblock \emph{\bibinfo{journal}{Science}} \textbf{\bibinfo{volume}{225}},
  \bibinfo{pages}{983--989} (\bibinfo{year}{1984}).

\bibitem{rev:goetze_leshouches1987}
\bibinfo{author}{G\"otze, W.}
\newblock \bibinfo{title}{Aspects of structural glass transitions}.
\newblock In \bibinfo{editor}{Hansen, J.~P.}, \bibinfo{editor}{Levesque, D.} \&
  \bibinfo{editor}{Zinn-Justin, J.} (eds.) \emph{\bibinfo{booktitle}{Liquids,
  freezing, and the glass transition}}, Proceedings of the LI Les Houches
  summer school (\bibinfo{publisher}{North-Holland}, \bibinfo{year}{1987}).

\bibitem{tension:imry75}
\bibinfo{author}{Imry, Y.} \& \bibinfo{author}{Ma, S.-k.}
\newblock \bibinfo{title}{Random-field instability of the ordered state of
  continuous symmetry}.
\newblock \emph{\bibinfo{journal}{Phys. Rev. Lett.}}
  \textbf{\bibinfo{volume}{35}}, \bibinfo{pages}{1399--1401}
  (\bibinfo{year}{1975}).

\bibitem{tension:halpin95}
\bibinfo{author}{Halpin-Healy, T.} \& \bibinfo{author}{Zhang, Y.-C.}
\newblock \bibinfo{title}{Kinetic roughening phenomena, stochastic growth,
  directed polymers and all that. {A}spects of multidisciplinary statistical
  mechanics}.
\newblock \emph{\bibinfo{journal}{Physics Reports}}
  \textbf{\bibinfo{volume}{254}}, \bibinfo{pages}{215--414}
  (\bibinfo{year}{1995}).

\bibitem{mosaic:xia00}
\bibinfo{author}{Xia, X.} \& \bibinfo{author}{Wolynes, P.~G.}
\newblock \bibinfo{title}{Fragilities of liquids predicted from the random
  first order transition theory of glasses}.
\newblock \emph{\bibinfo{journal}{Proc. Nac. Acad. Sci.}}
  \textbf{\bibinfo{volume}{97}}, \bibinfo{pages}{2990--2994}
  (\bibinfo{year}{2000}).

\bibitem{mosaic:dzero05}
\bibinfo{author}{Dzero, M.}, \bibinfo{author}{Schmalian, J.} \&
  \bibinfo{author}{Wolynes, P.~G.}
\newblock \bibinfo{title}{Activated events in glasses: The structure of
  entropic droplets}.
\newblock \emph{\bibinfo{journal}{Phys. Rev. B}} \textbf{\bibinfo{volume}{72}},
  \bibinfo{pages}{100201} (\bibinfo{year}{2005}).
\newblock \urlprefix\url{http://link.aps.org/abstract/PRB/v72/e100201}.

\bibitem{nucleation:franz05}
\bibinfo{author}{Franz, S.}
\newblock \bibinfo{title}{First steps of a nucleation theory in disordered
  systems}.
\newblock \emph{\bibinfo{journal}{J. Stat. Mech.}}
  \textbf{\bibinfo{volume}{2005}}, \bibinfo{pages}{P04001}
  (\bibinfo{year}{2005}).

\bibitem{tension:huse85}
\bibinfo{author}{Huse, D.~A.} \& \bibinfo{author}{Henley, C.~L.}
\newblock \bibinfo{title}{Pinning and roughening of domain walls in ising
  systems due to random impurities}.
\newblock \emph{\bibinfo{journal}{Phys. Rev. Lett.}}
  \textbf{\bibinfo{volume}{54}}, \bibinfo{pages}{2708--2711}
  (\bibinfo{year}{1985}).

\bibitem{self:prl02}
\bibinfo{author}{Grigera, T.~S.}, \bibinfo{author}{Cavagna, A.},
  \bibinfo{author}{Giardina, I.} \& \bibinfo{author}{Parisi, G.}
\newblock \bibinfo{title}{Geometric approach to the dynamic glass transition}.
\newblock \emph{\bibinfo{journal}{Phys. Rev. Lett.}}
  \textbf{\bibinfo{volume}{88}}, \bibinfo{pages}{055502}
  (\bibinfo{year}{2002}).
\newblock \urlprefix\url{http://dx.doi.org/10.1103/PhysRevLett.88.055502}.

\bibitem{self:jcp06}
\bibinfo{author}{Grigera, T.~S.}
\newblock \bibinfo{title}{Geometrical properties of the potential energy of the
  soft-sphere binary mixture}.
\newblock \emph{\bibinfo{journal}{J. Chem. Phys.}}
  \textbf{\bibinfo{volume}{124}}, \bibinfo{pages}{064502}
  (\bibinfo{year}{2006}).
\newblock \urlprefix\url{http://link.aip.org/link/?JCP/124/064502/1}.

\bibitem{self:coluzzi99}
\bibinfo{author}{Coluzzi, B.}, \bibinfo{author}{M\'{e}zard, M.},
  \bibinfo{author}{Parisi, G.} \& \bibinfo{author}{Verrocchio, P.}
\newblock \bibinfo{title}{Thermodynamics of binary mixture glasses}.
\newblock \emph{\bibinfo{journal}{The Journal of Chemical Physics}}
  \textbf{\bibinfo{volume}{111}}, \bibinfo{pages}{9039--9052}
  (\bibinfo{year}{1999}).
\newblock \urlprefix\url{http://link.aip.org/link/?JCP/111/9039/1}.

\bibitem{self:castellani05}
\bibinfo{author}{Castellani, T.} \& \bibinfo{author}{Cavagna, A.}
\newblock \bibinfo{title}{Spin-glass theory for pedestrians}.
\newblock \emph{\bibinfo{journal}{Journal of Statistical Mechanics: Theory and
  Experiment}} \textbf{\bibinfo{volume}{2005}}, \bibinfo{pages}{P05012}
  (\bibinfo{year}{2005}).
\newblock \urlprefix\url{http://stacks.iop.org/1742-5468/2005/P05012}.

\bibitem{soft-spheres:bernu87}
\bibinfo{author}{Bernu, B.}, \bibinfo{author}{Hansen, J.~P.},
  \bibinfo{author}{Hiwatari, Y.} \& \bibinfo{author}{Pastore, G.}
\newblock \bibinfo{title}{Soft-sphere model for the glass transition in binary
  alloys: Pair structure and self-diffusion}.
\newblock \emph{\bibinfo{journal}{Phys. Rev. A}} \textbf{\bibinfo{volume}{36}},
  \bibinfo{pages}{4891--4903} (\bibinfo{year}{1987}).

\bibitem{self:pre01}
\bibinfo{author}{Grigera, T.~S.} \& \bibinfo{author}{Parisi, G.}
\newblock \bibinfo{title}{Fast {M}onte {C}arlo algorithm for supercooled soft
  spheres}.
\newblock \emph{\bibinfo{journal}{Phys. Rev. E}} \textbf{\bibinfo{volume}{63}},
  \bibinfo{pages}{045102} (\bibinfo{year}{2001}).
\newblock \urlprefix\url{http://dx.doi.org/10.1103/PhysRevE.63.045102}.

\bibitem{soft-spheres:roux89}
\bibinfo{author}{Roux, J.-N.}, \bibinfo{author}{Barrat, J.-L.} \&
  \bibinfo{author}{Hansen, J.-P.}
\newblock \bibinfo{title}{Dynamical diagnostics for the glass transition in
  soft-sphere alloys}.
\newblock \emph{\bibinfo{journal}{J. Phys.: Condens. Matt.}}
  \textbf{\bibinfo{volume}{1}}, \bibinfo{pages}{7171--7186}
  (\bibinfo{year}{1989}).

\bibitem{algorithm:liu89}
\bibinfo{author}{Liu, D.~C.} \& \bibinfo{author}{Nocedal, J.}
\newblock \bibinfo{title}{On the limited memory {BFGS} method for large scale
  optimization}.
\newblock \emph{\bibinfo{journal}{Math. Programming}}
  \textbf{\bibinfo{volume}{45}}, \bibinfo{pages}{503--528}
  (\bibinfo{year}{1989}).

\end{thebibliography}

\end{document}